\documentclass[12pt]{iopart}
\pdfoutput=1
\usepackage{iopams}  
\usepackage{graphicx}
\usepackage{amssymb}
\usepackage{color}
\newcommand{\p}{\partial}

\newcommand{\bsf}[1]{\textsf{\textbf{#1}}}
\newcommand{\beq}{\begin{equation}}
\newcommand{\eeq}{\end{equation}}

\begin{document}

\title{Mesoscopic theory for fluctuating active nematics}

\author{Eric Bertin$^{1,2}$, Hugues Chat\'e$^{3,2,4}$, Francesco Ginelli$^{5}$, Shradha Mishra$^{6}$, Anton Peshkov$^{3,2,4}$, Sriram Ramaswamy$^{7,2}$\footnote{On leave from Department of Physics, Indian Institute of Science, Bangalore 560 012 India.}}

\address{$^1$ Universit\'e de Lyon, Laboratoire de Physique, ENS Lyon, CNRS, 46 All\'ee d'Italie, F-69007 Lyon}%

\address{$^2$ Max Planck Institute for the Physics of Complex Systems, N\"othnitzer Str. 38, 01187 Dresden, Germany}%

\address{$^3$ Service de Physique de l'Etat Condens\'e, CEA-Saclay, URA 2464 CNRS, 91191 Gif-sur-Yvette, France}

\address{$^4$ LPTMC, CNRS-UMR 7600, Universit\'e Pierre et Marie Curie, 75252 Paris, France}%

\address{$^5$ SUPA, Department of Physics and Institute for Complex Systems and Mathematical Biology, King's College, University of Aberdeen, Aberdeen AB24 3UE, United Kingdom}%

\address{$^6$ Department of Physics \& Meteorology, Indian Institute of Technology Kharagpur, Kharagpur 721 302, India}%

\address{$^7$ TIFR Centre for Interdisciplinary Sciences, Tata Institute of Fundamental Research, 21 Brundavan Colony, Narsingi, Hyderabad 500 075, India}

\date{\today}

\begin{abstract}
The term active nematics designates systems in which apolar
elongated particles spend energy to move randomly along their axis
and interact by inelastic collisions in the presence of noise.
Starting from a simple Vicsek-style model for active nematics, we
derive a mesoscopic theory, 
complete with effective multiplicative noise terms,
using a combination of kinetic theory and It\^o calculus approaches.
The stochastic partial differential equations thus obtained are shown to
recover the key terms argued in EPL \textbf{62} (2003) 196 to be at the
origin of anomalous number fluctuations and long-range correlations. Their
deterministic part is studied analytically, and is shown to give rise to
the long-wavelength instability at onset of nematic order
(see arXiv:1011.5408). The corresponding nonlinear density-segregated band solution
is given in a closed form.
\end{abstract}

\maketitle

\section{Introduction}

The study of collective properties of systems of interacting active
particles \cite{Ramaswamy0,ABP_review,MarchettiRMP} is currently 
attracting a great deal of interest. In active matter, particles extract energy from their surrounding and dissipate it to propel 
themselves in some coherent way in a viscous fluid and/or over a dissipative substrate. In this last case, or whenever
hydrodynamic effects can be neglected, physicists speak of ``dry active matter"
\cite{MarchettiRMP}.
Systems as diverse as animal flocks \cite{Parrish, Couzin, Ballerini}, human crowds \cite{Helbing1, Helbing2}, subcellular proteins \cite{MotilityAssays}, bacterial colonies \cite{Bacteria}, and driven granular matter \cite{Kudrolli, Ramaswamy1,VPD} have been described in this framework.

In the context of dry active matter, there is now some consensus in the physics community that minimal models 
such as the celebrated Vicsek model \cite{Vicsek,Chate1} play a crucial role, since they stand as simple representatives of 
universality classes which have started to emerge
from a combination of numerical and theoretical results: 
for instance, many different microscopic (particle) models have been shown to
exhibit the same collective properties as the Vicsek model, and the continuous equation proposed by Toner and Tu \cite{TT}
is widely believed to account for its collective properties.
Such hydrodynamic theories formulated at the mesoscopic level (stochastic
PDEs) are the natural framework to characterize and define universality classes. 

In early approaches these mesoscopic theories have been built on
the principle of including all that is not explicitly forbidden, retaining
all leading terms (in a gradient expansion sense) allowed by symmetries and
conservation laws
\cite{TT, Ramaswamy2003}. This grants access to the general structure of these equations
and has been successful in describing relevant features of active matter systems such as their anomalously large
number density fluctuations \cite{TT,Ramaswamy1, Swinney, VPD}.
Despite the attractions of a gradient expansion, it typically contains many
transport coefficients of unknown dependence on microscopic control parameters
and hydrodynamic fields such as local density. Moreover, the dependence of the noise terms on the dynamical
fields in such equations remains arbitrary, and frequently neglected, whereas
it could have profound consequences for important phenomena such as
spontaneous segregation, clustering and interface dynamics.   

Ideally, thus, one would be able to derive well-behaved mesoscopic theories using a systematic procedure 
starting from a given microscopic model. 
Kinetic-theory-like approaches \cite{Bertin2006, Bertin2009,
Ihle,Marchetti2,ShiMa} go one step
towards this goal, by allowing
one to compute hydrodynamic transport
coefficients and nonlinear terms. One of the most successful versions is
arguably the ``Boltzmann-Ginzburg-Landau" (BGL)
framework recently put forward by some of us \cite{Peshkov2012a,Peshkov2012b}, where, in the spirit of weakly nonlinear
analysis, one performs well-controlled expansions in the vicinity of ordering transitions.
Kinetic approaches alone thus yield good deterministic ``mean-field" equations but one still need to ``reintroduce"
fluctuations in order to get {\it bona fide} mesoscopic descriptions.

In this work, we show how this complete program can be achieved for the case of active nematics, 
i.e. systems where particles are energized individually but not really
self-{\em propelled}, moving along the axis of the
nematic degree of freedom they carry, with equal probability forward or back. 
(Think of shaken apolar rods aligning by inelastic collisions
\cite{Ramaswamy1}.)
Starting from the Vicsek-style model for active nematics introduced in \cite{Chate2},
we formulate a version of the BGL scheme mentioned above adapted to problems dominated by diffusion, 
derive  the corresponding hydrodynamic equations, and study their homogeneous and inhomogeneous solutions. 
In a last section, we show how these equations can be complemented by appropriate noise terms using a direct coarse-graining approach.

\section{Kinetic approach}
\label{sect-kinetic-theory}

\subsection{Microscopic dynamics}
\label{S2-1}

We consider the microscopic model for active nematics of \cite{Chate2} in two space dimensions.
This Vicsek-style model can be thought of as a minimal model for a
single layer of vibrated granular rods
\cite{Ramaswamy1} although it does not deal explicitly with any volume exclusion forces. 
Here, rather, pointwise particles $j=1,\ldots,N$ are characterized by their position $\bsf{x}^t_j$ and an 
axial direction $\theta^t_j \in [-\pi/2, \pi/2]$. They interact synchronously 
with all neighboring particles situated within distance $r_0$
in a characteristic driven-overdamped dynamics implemented at discrete timesteps $\Delta t$:
\begin{equation}
\theta^{t+\Delta t}_j = \frac{1}{2} \mathrm{Arg}\left[ \sum_{k \in V_j} e^{i 2
\theta^t_k} \right] + \psi^t_j
\label{eq:align}
\end{equation} 
\begin{equation}
\bsf{x}_j^{t+\Delta t} =  \bsf{x}_j^t + d_0 \,\kappa_j^t \, \hat{\bf n}_j^t\;,
\label{eq:stream}
\end{equation}
where $V_j$ is the neighborhood of particle $j$, $d_0<r_0$ is the elementary displacement,
$\hat{\bf n}^t_j \equiv \left(\cos \theta^t_j,\sin \theta^t_j\right)^T$ is the nematic director, and $\psi$ and $\kappa$
are two white noises: the random angle $\psi^t_j$, familiar of Vicsek-style models, is
drawn from a symmetric distribution $\tilde{P}_\eta(\psi)$ of variance
$\eta^2$, and the zero average bimodal noise $\kappa^t_j = \pm 1$ determines the actual orientation of motion.
Both noises are delta correlated, namely
$\langle \kappa^t_j \kappa^{t'}_k\rangle \sim \langle \psi^t_j \psi^{t'}_k\rangle \sim \delta_{t\,t'} \delta_{j\,k}$.

In the following, we adopt the convention
$\left[\hat{\bf n} \hat{\bf n}\right]_{\alpha \beta} \equiv \hat{n}_{\alpha} \hat{n}_{\beta}$ and label
coordinates by greek indices, $\alpha, \beta, \ldots=1,2$, summing over repeated indices. 

\subsection{Timescales and lengthscales}

We consider low density systems in which particles, at a given time, are either non-interacting or 
involved in a binary interaction. In this dilute limit we can neglect interactions between more than two
particles. We also treat interactions as collision-like events, with the mean intercollision time 
\begin{equation}
\tau_{\rm free}\approx \frac{\tau_d}{d_0^2 \,\rho_0}\,, 
\label{tfree}
\end{equation} 
where $\rho_0$ is the global particle density and $\tau_d$ is shortest microscopic timescale of the dynamics,
associated to the inversion of the rods direction of motion $\tau_d \sim \Delta t$. 
This inter-collision time is much larger than the collision timescale
\begin{equation}
\tau_{\rm coll} \approx \tau_d \left(\frac{r_0}{d_0}\right)^2\;.
\label{tcoll}
\end{equation}
For driven granular rods, $\tau_d$ may be thought of as the inverse of the shaking
frequency, and for typical parameters it is much smaller 
than both the collision ($\tau_{\rm coll}$) and the mean intercollision ($\tau_{\rm free}$) timescales; at low enough densities 
$\tau_d \ll \tau_{\rm coll} \ll \tau_{\rm free}$.
Note that the timescales (\ref{tcoll})-(\ref{tfree})
are different from the ones characteristic of ballistic dynamics \cite{Bertin2009}.

To develop a kinetic approach we consider a mesoscopic
timescale $\tau_B$ such that $\tau_{\rm coll} \ll \tau_B \ll \tau_{\rm free}$. 
As a consequence, we will treat the inversion of the
direction of motion as a noisy term through It\^o stochastic calculus
\cite{Ito}.
We also consider a mesoscopic coarse-graining lengthscale  $\ell_B$ which, while being much smaller than the system size $L$,
is larger than the microscopic scales, such as the step-size $d_0$,
the mean interparticle distance $\rho_0^{-1/2}$ and the interaction range $r_0$.
To summarize, in a dilute system one has 
\begin{equation}
\tau_d \ll \tau_d \left(\frac{r_0}{d_0}\right)^2 \ll \tau_B \ll \frac{\tau_d}{d_0^2 \rho_0}
\label{eq:time}
\end{equation}
and 
\begin{equation}
d_0 < r_0 \ll \frac{1}{\sqrt{\rho_0}} \ll \ell_B \ll L 
\label{eq:length}
\end{equation}
where $L$ is the system size and we have made explicit the condition that the typical coarse-graining lengthscale
$\ell_B$ is such that many particles are contained in a box of linear size 
$\ell_B$, that is $\rho_0 \ell_B^2  \gg 1$.

\subsection{Master equation}

We now write down a Boltzmann-like master equation in terms of the single particle probability distribution
$f(\bsf{x}, \theta, t)$, with $-\frac{\pi}{2} < \theta \le \frac{\pi}{2}$,
evolving over the timescale $\Delta t \approx \tau_B$. 
The minimal spatial resolution is such that many particles are contained in a spatial volume $d^2x$
centered around the position $\bsf{x}$. Moreover, we consider a dilute system, so that interactions (collisions) 
between particles are sufficiently rare to justify 
i) {\it binary interactions} (as explained above, particles then either self-diffuse or experience noisy binary, 
collision-like interactions), ii) decorrelation of the orientation
between successive binary collisions of the same pair of particles,
that is 
$f_2(\bsf{x}, \theta_1, \theta_2, t) \approx f(\bsf{x}, \theta_1, t)f(\bsf{x}, \theta_2, t)$.

We first omit collisions and angular diffusion, only considering Eq.~(\ref{eq:stream}) to get
\begin{equation}
f(\bsf{x}, \theta, t+\Delta t) = \frac{1}{2} \left[f(\bsf{x}+\hat{\bf n}(\theta)  d_0 , \theta, t)+ 
f(\bsf{x}-\hat{\bf n}(\theta) d_0 , \theta, t) \right]\;,
\end{equation}
where we have considered that a particle moves along one of the two orientations of $\hat{\bf n}$ with equal probability.
On the mesoscopic timescale $\tau_B\gg\tau_d \sim \Delta t$,
It\^o calculus \cite{Ito} to second order gives
\begin{equation}
\partial_t f(\bsf{x}, \theta, t) =
D_0 \partial_{\alpha}\partial_{\beta} [ \hat{n}_{\alpha}(\theta)
\hat{n}_{\beta}(\theta)   f(\bsf{x}, \theta, t) ]
\label{B1}
\end{equation}
where 
\beq
D_0=\frac{d_0^2}{2\tau_d}
\label{eq:D}
\eeq
is the microscopic diffusion parameter. \\

To account for angular diffusion and binary collisions,
the appropriate integrals need to be added to the right hand side of Eq.~(\ref{B1}),
\begin{equation}
\partial_t f(\bsf{x}, \theta, t) =
D_0 \, \partial_{\alpha}\partial_{\beta} [ \hat{n}_{\alpha}(\theta) \hat{n}_{\beta}(\theta)
f(\bsf{x}, \theta, t) ] +I_{\rm diff}[f]+I_{\rm coll}[f,f]\,.
\label{B2}
\end{equation}
The diffusion integral describes self-diffusion which takes
place at a rate $\lambda = 1/\tau_d$
\begin{equation}
I_{\rm diff} [f] = -\lambda f(\theta) 
+ \lambda \int_{-\pi/2}^{\pi/2} d \theta' f(\theta') 
\int_{-\infty}^{\infty} d \zeta P(\zeta) \,\delta_{\pi} (\theta' - \theta + \zeta)\nonumber
\label{Idiff}
\end{equation}
where we used the simplified notation $f(\theta) \equiv f(\bsf{x}, \theta, t)$,
$\delta_{\pi}$ is a generalized Dirac delta imposing that the argument is equal to zero modulo $\pi$ and
$P(\zeta)$ is a symmetric noise distribution of variance $\sigma^2$, corresponding to the effective noise arising at the timescale $\tau_B$
from the sum of the microscopic stochastic contributions to angular dynamics.

Binary collisions are described by
\begin{eqnarray}
 \hspace{-1.5 cm} I_{\rm coll} [f,f] & = & -f(\theta) \int_{-\pi/2}^{\pi/2} \!\!\! d \theta' f(\theta') 
K(\theta,\theta') \\
&+& \int_{-\pi/2}^{\pi/2} \!\!\!d \theta_1 \!\int_{-\pi/2}^{\pi/2} \!\!\!d \theta_2
f(\theta_1) K(\theta_1, \theta_2) f(\theta_2)  
\int_{-\infty}^{\infty} \!\!d \zeta P(\zeta) \,
\delta_{\pi} (\Psi(\theta_1,\theta_2) \!-\! \theta\! +\!\zeta)\,,\nonumber
\label{Icoll}
\end{eqnarray}
where, for the sake of simplicity, we have used the same noise distribution
$P(\zeta)$
as in the self-diffusion integral, and
the out-coming angle $\Psi$ from deterministic binary collisions is,
for $-\frac{\pi}{2} < \theta_1, \theta_2 \le \frac{\pi}{2}$,
\begin{equation}
\label{def:def-Phi}
\Psi(\theta_1,\theta_2) = \frac{1}{2}(\theta_1 + \theta_2) + h(\theta_1 - \theta_2)
\;\; {\rm with} \;\;
h(\theta) =\left\{
\begin{array}{lr}
0 & \mathrm{if}\;\;|\theta| \leq \frac{\pi}{2} \\
\frac{\pi}{2} &
\mathrm{if}\;\; \frac{\pi}{2} < |\theta| \leq \pi
\end{array}
\right.
\end{equation}
Note that the role of the function $h(\theta)$ is to ensure that
$\Psi(\theta_1,\theta_2)$ is $\pi$-periodic
with respect to $\theta_1$ and $\theta_2$ independently.
The collision kernel $K(\theta_1,\theta_2)$, {\it i.e.} the number
of collisions per unit time and volume, is calculated as follows. Consider two particles with
nematic axis $\hat{\bf n}(\theta)$ and $\hat{\bf n}(\theta')$ located in the
volume $d^2x$ centered around position $\bsf{x}$. 
In the reference frame of the first particle the second one diffuses
either along the $|\hat{\bf n}(\theta) - \hat{\bf n}(\theta')|$ or the 
$|\hat{\bf n}(\theta) + \hat{\bf n}(\theta')|$ nematic axis. In unit time,
taking into account the characteristic timescales $\tau_d$ and step-size $d_0$
of its motion, it sweeps a surface (its cross section, which is conserved going back
to the lab reference frame) equal to 
\begin{eqnarray}
K(\theta,\theta') &=& \frac{r_0 d_0}{\tau_d} \left[ |\hat{\bf n}(\theta) - \hat{\bf n}(\theta')| +
|\hat{\bf n}(\theta) + \hat{\bf n}(\theta')|\right] \nonumber\\
&=& 2 \alpha_0 
\left[\left|\sin \frac{\theta - \theta'}{2}\right|
+\left|\cos \frac{\theta - \theta'}{2}\right|\right]\,,
\label{collisionkernel}
\end{eqnarray}
where we have introduced the microscopic collision parameter
\beq
\alpha_0 = \frac{r_0 d_0}{\tau_d}\,.
\label{eq:alpha0}
\eeq
Note that $K(\theta,\theta')\equiv \tilde{K}(\theta-\theta')$ is an even function of the difference $(\theta-\theta')$, 
and fulfills the nematic symmetry, being invariant under rotation of either angle by $\pi$.

Before proceeding to derive hydrodynamic equations, we simplify all notations by rescaling time
$\tilde{t}=\lambda t=t/\tau_d$
and space $\tilde{x}=\frac{\sqrt{2}}{d_{0}}x$. As in \cite{Peshkov2012a,Peshkov2012b} 
we also set the collision surface $S=2r_0d_0$ to $1$ by a global rescaling of the one-particle probability density $f$, without loss of generality. 
This amounts to set $\lambda_0=1$, $D_0=1$ and $2\alpha_0=1$, so that, dropping the tildes,
our Boltzmann-like master equation now depends only on the global density
$\rho_0$ and the noise intensity $\sigma$.

\subsection{Hydrodynamic description}
\label{sect-hydro}

In two spatial dimensions, hydrodynamic fields can be obtained by 
expanding the single particle probability density $f$
in Fourier series of its angular variable $\theta \in
[-\pi/2,\pi/2]$\footnote{These $k$-modes are
equivalent to even harmonics if one would define particles orientation in
$[-\pi,\pi]$ in spite of the symmetry under rotations by $\pi$ (with
odd ones being zero by symmetry).}:
\begin{equation}
\label{expansion}
f(\bsf{x}, \theta, t) = \frac{1}{\pi} \sum_{k=-\infty}^{k=\infty} \hat{f}_k(\bsf{x}, t) e^{-i 2 k \theta}
\end{equation}
and
\begin{equation}
\label{expansion2}
\hat{f}_k(\bsf{x}, t) = \int_{-\pi/2}^{\pi/2} d \theta 
f(\bsf{x}, \theta, t) e^{i 2 k \theta} \;.
\end{equation}
The number density and the density-weighted nematic tensor field ${\bf w}\equiv \rho {\bf Q}$ are then given by
\begin{equation}
\rho(\bsf{x}, t) = \int_{-\pi/2}^{\pi/2} d \theta 
f(\bsf{x}, \theta, t) = \hat{f}_0(\bsf{x}, t)
\end{equation}
and
\begin{eqnarray}
w_{11}(\bsf{x}, t) = - w_{22}(\bsf{x}, t) &=& 
\frac{1}{2}\int_{-\pi/2}^{\pi/2} d \theta 
f(\bsf{x}, \theta, t) \cos (2 \theta) = \frac{1}{2} {\rm Re} \hat{f}_1(\bsf{x}, t) 
\label{Qfield1}\\
w_{12}(\bsf{x}, t) = w_{21}(\bsf{x}, t) &=& 
\frac{1}{2}\int_{-\pi/2}^{\pi/2} d \theta 
f(\bsf{x}, \theta, t) \sin (2 \theta) = \frac{1}{2} {\rm Im} \hat{f}_1({\bf
  x}, t)
\label{Qfield2}
\end{eqnarray}
Note that when ${\rm Im} \hat{f}_1=0$ the nematic field is aligned either
along the $x$ (${\rm Re} \hat{f}_1>0$) or the $y$ (${\rm Re} \hat{f}_1<0$) axis.

Injecting the Fourier expansion (\ref{expansion}) in the master equation (\ref{B2}), one gets, after some 
lengthy calculations detailed in \ref{App-Fourier-MEq},
the infinite hierarchy:
\begin{eqnarray}
\p_t \hat{f}_k(\bsf{x}, t) &=& \frac{1}{2} \Delta \hat{f}_k(\bsf{x}, t) 
+  \frac{1}{4}\left( \nabla^{*2} \hat{f}_{k+1} + \nabla^{2} \hat{f}_{k-1} \right)
+  \left[ \hat{P}_k -1 \right]\hat{f}_k (\bsf{x}, t)\nonumber\\ 
&+& \frac{1}{\pi} \sum_q \hat{f}_q (\bsf{x}, t) \hat{f}_{k-q} (\bsf{x}, t)
\left[\hat{P}_k \hat{J}_{k,q} -\frac{4}{1-16q^2}\right]
\label{Hydro0}
\end{eqnarray}
where $\hat{P}_k$ is the Fourier transform of
the noise distribution $P(\zeta)$ (namely, $\hat{P}_k=\int_{-\infty}^{\infty} d\zeta P(\zeta) e^{i2k\zeta}$) and
\beq
\hat{J}_{k,q}
=4\,\frac{1+2\sqrt{2}(2q-k)(-1)^q\sin\left(\frac{k\pi}{2}\right)}{1-4(2q-k)^2}
\eeq
and we have introduced the following ``complex" operators
\begin{eqnarray*}
\nabla & \equiv & \partial_{x}+{\rm i}\partial_{y}\\
\nabla^* & \equiv & \partial_{x}-{\rm i}\partial_{y}\\
\Delta & \equiv & \nabla \nabla^*\\
\nabla^{2} & \equiv & \nabla\nabla\\
{\nabla^{*}}^2 & \equiv & \nabla^* \nabla^*
\end{eqnarray*}
The equation at order $k=0$ is thus expressed in the simple form
\begin{equation}
\partial_{t}\rho =  \frac{1}{2}\Delta\rho+\frac{1}{2}{\rm Re}\left({\nabla^*}^{2}\hat{f}_{1}\right)
\label{eq:RhoComplex}
\end{equation}
and is nothing but the continuity equation for diffusive active matter with
local anisotropy characterized by $\hat{f}_1$.

Eq.~(\ref{Hydro0}) possesses a trivial, isotropic  and homogeneous solution: 
$\rho(\bsf{x}, t) = \hat{f}_0(\bsf{x}, t) = \rho_0$ and $\hat{f}_k(\bsf{x}, t)=0$ for $|k|>0$.
We are interested in a nematically ordered homogeneous solution which
could eventually 
arise following some instability of the isotropic solution above.
In analogy to the scaling ansatz used for polar particles \cite{Bertin2009,Peshkov2012b}, the interaction term in Eq.~(\ref{Hydro0}) suggests
a simple scaling ansatz to close the infinite hierarchy of equations on $\hat{f}_{k}(\bsf{x}, t)$:
Near an instability threshold with continuous onset, Fourier coefficients
should scale as
$\hat{f}_{k}(\bsf{x}, t) \sim \epsilon^{|k|}$
where $\epsilon$ is a small parameter characterizing the distance to threshold. 
Moreover, the curvature induced current (last term of (\ref{eq:RhoComplex})) also induces 
an order $\epsilon$ variation in the density field, $\rho(\bsf{x}, t)
- \rho_0 \sim \epsilon$.
Then, assuming spatial derivatives to be of order $\epsilon$, the request that all terms in Eq.~(\ref{eq:RhoComplex}) are of the same order also fixes the diffusive
structure of the scaling of time and spatial gradients: $\p_t \sim \nabla^2 \sim \Delta \sim \epsilon^2$.

Using the above scaling ansatz, we proceed by discarding all terms appearing in (\ref{Hydro0}) of order higher than $\epsilon^3$. 
For $k=1,2$ we get:
\begin{equation}
\p_t \hat{f}_1 = \frac{1}{2} \Delta \hat{f}_1 
+ \frac{1}{4} \nabla^2  \rho
+a_1(\rho) \hat{f}_1 + b_1  \hat{f}_1^* \hat{f}_2
\label{eq:k1}
\end{equation}
and
\begin{equation}
0=
 \frac{1}{4} \nabla^2 \hat{f}_1
- a_2(\rho) \hat{f}_2 + b_2 \hat{f}_1 \hat{f}_1 
\label{eq:k2}
\end{equation}
where the coefficients are
\begin{equation}
a_1(\rho)=\frac{8}{3\pi}\left[(2\sqrt{2}-1)\hat{P}_{1}-\frac{7}{5}\right]\rho-(1-\hat{P}_{1})\,,
\end{equation}
\begin{equation}
b_1=\frac{8}{315\pi} \left[13-9 \hat{P}_1 (1+6 \sqrt{2} )\right]
\end{equation}
\begin{equation}
a_2(\rho)=(1-\hat{P}_2)
+\frac{8}{3\pi}\left(\frac{\hat{P}_2}{5}+\frac{31}{21}\right) \rho
\end{equation}
and
\begin{equation}
b_2=\frac{4}{\pi} \left(\frac{1}{15} +\hat{P}_2 \right)\,.
\end{equation}
Eq.~(\ref{eq:k2}) shows that at this order $\hat{f}_2$ is enslaved to $\hat{f}_1$ (given that $a_2>0$) and, further,
\begin{equation} \label{eq-f2-closure}
a_2(\rho_0)\hat{f}_2 \approx \frac{1}{4} \nabla^2 \hat{f}_1
+b_2\hat{f}_1\hat{f}_1\,,
\end{equation}
where the coefficient $a_2$ is evaluated at the mean density $\rho_0$,
since the $\delta \rho = \rho-\rho_0$ corrections are of higher order.
By substituting Eq.~(\ref{eq:k2}) into (\ref{eq:k1}) one finally gets,
neglecting the term $\hat{f}_1^* \nabla^2 \hat{f}_1 \sim \epsilon^4$,
\begin{equation}
\partial_{t}\hat{f}_{1} = \left(\mu-\xi\left|\hat{f}_{1}\right|^{2}\right)\hat{f}_{1}+\frac{1}{4}\nabla^{2}\rho+\frac{1}{2}\Delta \hat{f}_{1} 
\label{eq:last}
\end{equation}
where we have introduced the transport coefficients
\begin{eqnarray}
&\mu & =  \frac{8}{3\pi}\left[\left(2\sqrt{2}-1\right)\hat{P}_{1}-\frac{7}{5}\right]\rho-\left(1-\hat{P}_{1}\right)
\label{eq:mu} \\
&\xi & =  \frac{32 \nu}{35\pi^{2}}\left[\frac{1}{15}+\hat{P}_{2}\right]
\left[\left(1+6\sqrt{2}\right)\hat{P}_{1}-\frac{13}{9}\right]
\label{eq:xi} \\
{\rm with}\;\; & \nu & =  \left[\frac{8}{3\pi}\left(\frac{31}{21}+\frac{\hat{P}_{2}}{5}\right)\rho_0+\left(1-\hat{P}_{2}\right)\right]^{-1} \;.
\label{eq:nu}
\end{eqnarray}
Note that the coefficient $\xi$ is only a function of the average density
$\rho_0$, as space and time dependent corrections are of order $\epsilon^4$.
Note also that the coefficients $\mu$ and $\xi$ are exactly
the same as those found for the nematic field equation of nematically-aligning polar particles \cite{Peshkov2012b} 
\footnote{Note that in \cite{Peshkov2012b}, the equations obtained are
not entirely correct: (i) there is a sign error and a misplaced factor
$\pi$ in the expression of $\xi$; (ii) the term $\frac{\nu}{4}\nabla^2
f_2$ should read $\frac{\nu}{4}\Delta f_2$, where $\Delta$ is the
Laplacian. In addition, let us emphasize that the Fourier coefficients
$\hat{P}_k$ have a different definition in \cite{Peshkov2012b}, due to
the absence of global nematic symmetry: $\hat{P}_k$ here corresponds to
$\hat{P}_{2k}$ in \cite{Peshkov2012b}, leading to (only apparent)
differences.}

 Eqs.~(\ref{eq:RhoComplex}) and (\ref{eq:last}) can be expressed in tensorial notation.
To this aim, we introduce the linear differential operator ${\bf \Gamma}$, such that
 $\Gamma_{11}=-\Gamma_{22}\equiv \p_1\p_1-\p_2\p_2$ and
 $\Gamma_{12}=\Gamma_{21} \equiv 2\p_1\p_2$,
and the Frobenius inner product 
${\bf A}:{\bf B}= A_{\alpha\beta} B_{\alpha\beta}$
(note that ${\bf w}:{\bf w}=||{\bf w}||^2$ and ${\bf \Gamma}:{\bf w}=2 \p_\alpha\p_\beta w_{\alpha\beta}$).
After some 
manipulation of the terms and the use of Eqs.~(\ref{Qfield1},\ref{Qfield2}), 
we obtain the hydrodynamic equations for the density and nematic field
\begin{eqnarray}
\p_t \rho &=& \frac{1}{2} \Delta \rho 
+ \frac{1}{2} ( {\bf \Gamma} : {\bf w} )\,,
\label{eq:rho2} \\
\p_t {\bf w} &=& \mu \,{\bf w} - 2\xi  \,{\bf w} \left({\bf w}:{\bf w}\right) 
+ \frac{1}{2} \Delta {\bf w}  + \frac{1}{8}{\bf \Gamma} \rho
\label{eq:Q2}
\end{eqnarray}

Although the tensorial notation might be more familiar to some readers, it is in fact easier here 
to continue manipulating the complex field $\hat{f}_1$ and the complex operators defined above. 
Moreover, in the following we drop the ``$\,\hat{\;}\,$'' superscript to ease
notations. Eqs. (\ref{eq:rho2}) and (\ref{eq:Q2}) are also derived from
an apolar Vicsek-style model in \cite{shradha_thesis}.

The parameter-free character of the Laplacian term 
in (\ref{eq:Q2}) means, consistent with our expansion in $\epsilon$, that the
nematic phase of our system will be characterized by a single Frank constant
\cite{Frank}. The nonlinearities studied in \cite{MishraSimha} are therefore
also absent to this order.
The last term in Eq.~(\ref{eq:RhoComplex}) (or Eq.~(\ref{eq:rho2})), {\it i.e.} 
$\frac{1}{2}{\rm Re}\left(\nabla^{2}{f}_{1}\right)$ (or $\frac{1}{2} ( {\bf \Gamma} : {\bf w} )$),
is a curvature induced current which couples the density
and the nematic field. While its existence was first deduced from general
principles \cite{Ramaswamy2003},
here we have computed it directly from microscopic dynamics.
Our calculations also give an exact expression for the corresponding
transport coefficient, which is equal to the diffusive one (in
Eq.~(\ref{eq:RhoComplex}) or Eq.~(\ref{eq:rho2})), 
here set to $1/2$ by our rescaling.
In \ref{app-curv-current}, we show explicitly that this
curvature-induced current originates from the coupling of orientation
with motility.

We note finally that Eqs.~(\ref{eq:rho2},\ref{eq:Q2}) are similar to those
found by Baskaran and Marchetti \cite{Marchetti2} but simpler, largely due to
our simpler starting point.

\subsection{Homogeneous solutions} 

From now on, we use for $P(\zeta)$ a centered Gaussian distribution
of variance $\sigma^2$, in which case $\hat{P}_k = e^{-2k^2\sigma^2}$.
The linear stability with respect to homogeneous perturbations of the disordered  solution 
$\rho (\bsf{x}, t)=\rho_0$, $\hat{f}_1 (\bsf{x}, t)=0$
is given by the sign of $\mu(\rho_0)$ which yields the basic transition line
\beq
\sigma_{\rm t} = \sqrt{\frac{1}{2}\ln \left[5\, \frac{8 (2 \sqrt{2}
      -1)\rho_0 +3 \pi}{56 \rho_0 + 15 \pi}\right]} \;.
\label{eq:sigmac}
\eeq
Note that in the dilute limit $\rho_0 \ll1$, where the equations have been derived, 
one has $\sigma_{\rm t} \sim \sqrt{\rho_0}$.

For $\sigma<\sigma_{\rm t}$,
$\mu>0$, and the homogeneous nematically ordered solution
\beq
|f_1|=\sqrt{\frac{\mu}{\xi}}
\eeq
exists and is stable w.r.t.~homogeneous perturbations. The critical line
is shown in Fig.~\ref{fig:stab_diag}a (black solid line).
Note that for $\sigma < \sigma_{\rm t}$, all transport coefficients
(\ref{eq:mu}-\ref{eq:nu}) are positive.
This will be useful in the rest of the paper.

\section{Linear stability analysis}

We now study the linear stability of the above homogeneous solutions w.r.t.~to arbitrary perturbations.
Linearizing Eqs.~(\ref{eq:RhoComplex}) and (\ref{eq:last}) around a homogeneous solution,
$f_1=f_{1,0} +\delta f_1$ and $\rho=\rho_0 + \delta \rho$, one has
\begin{eqnarray}
\label{lin:rho}
\hspace{-2 cm}
\p_t \delta \rho &=& \frac{1}{2} \Delta \delta \rho
+ \frac{1}{2} {\rm Re}\left({\nabla^*}^2 \delta f_1\right)\\
\hspace{-2 cm}
\label{lin:Q}
\partial_{t}\delta f_{1} &=&  \left(\mu_0-\xi\left|f_{1,0}\right|^{2}\right) \delta f_{1} + \mu' f_{1,0} \, \delta\rho
- 2\xi f_{1,0}\, \mathbb{\rm Re}\left({f}_{1,0}^*\, \delta f_{1}\right)
+ \frac{1}{4}\nabla^{2}\delta\rho+\frac{1}{2}\Delta\delta f_{1}
\end{eqnarray}
where $\mu_0 \equiv \mu(\rho_0)$ and
$\mu'$ is the derivative of $\mu$ w.r.t. $\rho$.
We then introduce the real and imaginary parts of the order parameter
perturbation, $\delta f_1 = \delta f_1^{(R)} + i\delta f_1^{(I)}$, and express
the spatial dependence of all perturbation fields in Fourier space,
with a wavevector ${\bf q} = (q_x,q_y)$, by introducing the ansatz
\begin{eqnarray}
\label{eq:ansatz2a}
\delta \rho(\bsf{x}, t) &=& \delta \rho_{\bf q} \,e^{s t + i {\bf q}{\bf r}} \;,\\
\delta f_1^{(R)} (\bsf{x}, t) &=& \delta f_{1,{\bf q}}^{(R)}\, e^{s t + i {\bf q}{\bf r}} \;, \quad
\delta f_1^{(I)} (\bsf{x}, t) = \delta f_{1,{\bf q}}^{(I)}\, e^{s t + i {\bf q}{\bf r}} \;.
\label{eq:ansatz2b}
\end{eqnarray}
The stability of the stationary solution $f_{1,0}$ is then ruled by
the real part of the growth rate $s$.

\subsection{Stability of the disordered isotropic solution}

We first study the stability of the disordered solution $f_{1,0}=0$,
in the case $\mu_0<0$.
Substituting Eqs.~(\ref{eq:ansatz2a}), (\ref{eq:ansatz2b}) in Eqs.~(\ref{lin:rho}), (\ref{lin:Q}), one has
\begin{eqnarray}
\label{eq:lin}
s\, \delta \rho_{\bf q} &=& -\frac{q^2}{2} \,\delta \rho_{\bf q}
-\frac{1}{2} (q_x^2-q_y^2) \delta f_{1,\bf q}^{(R)}
-q_x q_y \delta f_{1,\bf q}^{(I)} \,,\\
s\, \delta f_{1,\bf q}^{(R)} &=&  - \frac{1}{4} (q_x^2-q_y^2) \delta \rho_{\bf q}
+ \left( \mu_0 -\frac{q^2}{2}\right) \delta f_{1,\bf q}^{(R)} \,, \nonumber\\
s\, \delta f_{1,\bf q}^{(I)} &=&  -\frac{1}{2} q_x q_y \, \delta \rho_{\bf q} + \left( \mu_0 -\frac{q^2}{2}\right) \delta f_{1,\bf q}^{(I)} \,, \nonumber
\end{eqnarray}
where $q^2 = q_x^2 + q_y^2$.
All directions of the wavevector ${\bf q}$ being equivalent,
we choose for simplicity $q_x=q$ and $q_y=0$.
From Eq.~(\ref{eq:lin}), one then sees that the component $\delta f_{1,\bf q}^{(I)}$
becomes independent from $\delta \rho_{\bf q}$ and $\delta f_{1,\bf q}^{(R)}$,
yielding the negative eigenvalue $s=\mu_0 -\frac{q^2}{2}$.
The eigenvalues of the remaining $2\times 2$ block of the stability matrix
are solutions of the second order polynomial
\beq
s^2+s\left[\,q^2 - \mu_0 \right] 
+\frac{q^2}{2} \left[\frac{q^2}{4} - \mu_0  \right] \equiv s^2+\beta_1 s + \beta_0 = 0 \;.
\eeq
In the disordered state $\mu_0<0$, so that
$\beta_1$ and $\beta_2$ are positive and one always has ${\rm Re} (s)<0$.
Therefore, the homogeneous disordered solution
is stable w.r.t.~to all perturbations if $\mu_0<0$, {\it i.e.} $\sigma>\sigma_{\rm t}$.

\subsection{Stability of the ordered solution}

\begin{figure}
\includegraphics[width=0.8\textwidth]{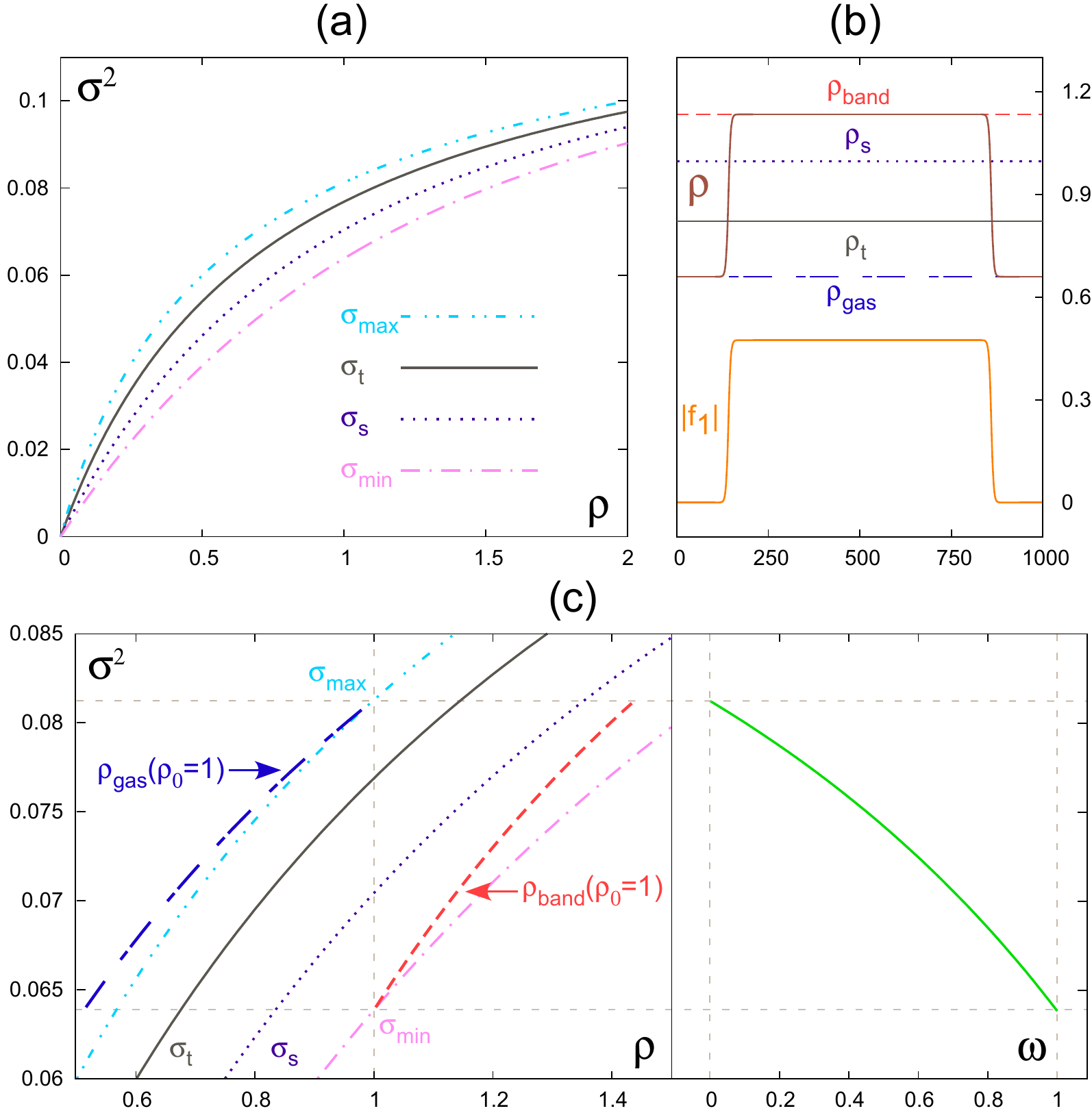}
\caption{(a) Basic stability diagram. The line $\sigma_t$ (solid, black) marks the linear instability of the disordered homogeneous solution. The ordered homogeneous solution is linearly unstable to large wavelengths between the $\sigma_t$ and $\sigma_s$ (dotted, purple) lines, and linearly stable below the 
$\sigma_s$ line. The $\sigma_{\rm min}$ and $\sigma_{\rm max}$ lines mark the domain of existence of the 
band solution (\ref{eq:f1-band}).
(b) Density and order profile of the band solution for $\rho_0=1$, $\sigma=0.265$, $L=1000$; note that the lower and upper levels ($\rho_{\rm gas}$ and $\rho_{\rm band}$) are respectively lower than $\rho_t$ and higher than $\rho_s$, {\it i.e.} such that the corresponding homogeneous solution are lineally stable.
(c): properties of the band solutions for $\rho_0=1$: left: values of $\rho_{\rm gas}$ (long dash, dark blue line) and $\rho_{\rm band}$ (dashed, red line) as 
$\sigma$ varies between $\sigma_{\rm min}$ and $\sigma_{\rm max}$; right: corresponding variation of the surface fraction $\omega$.
}
\label{fig:stab_diag}
\end{figure}

To study the stability of the anisotropic ordered solution, it is convenient to
choose a reference frame in which order is along one of the axes:
\beq
{\rm Re} \left( f_{1,0} \right) =\pm \sqrt{\frac{\mu_0}{\xi}} 
\; , \quad {\rm Im} \left( f_{1,0} \right) =0 \;.
\label{ref-frame}
\eeq
This solution is aligned along $x$, if $f_{1,0}$ is positive, or along $y$ if negative. 
For simplicity we will concentrate further on the case $f_{1,0}\geq 0$, i.e.,
on the nematic solution aligned along the $x$ axis.
The real part $\delta f_1^{(R)}$ of the nematic field perturbation 
describes changes in the modulus $|f_{1,0}|$, 
and the imaginary part $\delta f_1^{(I)}$ describes perturbations
perpendicular to the nematic orientation.
The ansatz (\ref{eq:ansatz2a}), (\ref{eq:ansatz2b}) then yields the
three coupled linear equations
\begin{eqnarray}
\label{eq:lin2}
s\, \delta \rho_{\bf q} &=& -\frac{q^2}{2} \,\delta \rho_{\bf q}
-\frac{1}{2} (q_x^2-q_y^2) \delta f_{1,\bf q}^{(R)}
-q_x q_y\, \delta f_{1,\bf q}^{(I)} \,,\\
s\, \delta f_{1,\bf q}^{(R)} &=& \left[ \mu' f_{1,0} -
  \frac{1}{4} (q_x^2-q_y^2) \right]\delta \rho_{\bf q} - \left[2 \mu_0 
+ \frac{q^2}{2} \right]  \delta f_{1,\bf q}^{(R)} \,, \nonumber\\
s\, \delta f_{1,\bf q}^{(I)} &=& - \frac{1}{2} q_x q_y\, \delta \rho_{\bf q} 
- \frac{q^2}{2} \delta f_{1,\bf q}^{(I)} \,. \nonumber
\end{eqnarray}
We performed a full numerical stability analysis of these equations. 
The results are presented in Fig.~\ref{fig:stab_diag}.
The transition to the homogeneous solution is given by the line $\sigma_t$. 
This solution is unstable to finite wavelength transversal
perturbations of angle $|\theta|>\frac{\pi}{4}$ between the lines $\sigma_t$ and $\sigma_s$ (dotted purple line in Fig.~\ref{fig:stab_diag}), but is 
stable deeper in the ordered phase.

Two remarks are in order.
First, the angle of the most unstable mode is here always perfectly $\frac{\pi}{2}$. It is thus possible 
to obtain the ``restabilization'' line $\sigma_s$
analytically as shown below. Second, there is no spurious instability at low noise and/or high density
(although we have found that such an instability appears if the truncation of the equations is made to the fourth order). 

To obtain the analytic expression of the line $\sigma_s$,
we write the wavevector in terms of its modulus $q$ and its 
angle $\theta_{\bf q}$, so that
$q_x^2-q_y^2 = q^2 \cos 2\theta_{\bf q}$ and
$2q_xq_y =q^2 \sin 2\theta_{\bf q}$.
We can then analyze Eqs.~(\ref{eq:lin2}) in the longitudinal and perpendicular 
wavedirections $\theta_{\bf q}=0\,,\pm
\frac{\pi}{2}$, where the imaginary perturbation $\delta f_{1,\bf q}^{(I)}$ decouples from the other two. 
The latter is stable towards long wavelength perturbations, since the
corresponding eigenvalue $s=-q^2/2$ is negative.
The stability towards density and real perturbations depends on a $2\times2$ matrix 
which yields the quadratic eigenvalue equation
\beq
s^2 +\left[ 2\mu_0 +q^2 \right] s + 
\left[ \left(\frac{\pm\mu' f_{1,0}}{2} +\mu_0 \right)q^2+\frac{q^4}{8} \right] = 0
\label{eq:stab_solution0}
\eeq
whose solutions are
\begin{equation}
s=\frac{1}{2} \left[ -2\mu_0-q^2 \pm \sqrt{4\mu_0^2 \mp 2\mu'f_{1,0}q^2+\frac{q^4}{2}} \right] \;.
\label{eq:stab_solution}
\end{equation}
The sign $\pm$ in front of the $\mu'f_{1,0}$ term in Eq.~(\ref{eq:stab_solution0})
corresponds to the case $\theta_{\bf q}=0$ (positive sign)
and $\theta_{\bf q}=\frac{\pi}{2}$ (negative sign) respectively.
Note that $\mu'$ is strictly positive, as typical for all active matter system with
metric interactions, where the interaction rate grows with local
density. Also $\mu_0$ is positive and of order
$\epsilon^2$ (see Eq.~(\ref{ref-frame})). 
It it thus easy to see that in the case of large $q$,  $\Re [s]\leq0$.
For small values of $q$, we perform an expansion to order $q^2$
of the largest growth rate $s_+$, obtained by taking the positive sign in front of the square root
in Eq.~(\ref{eq:stab_solution}), leading to
\beq
s_+=\frac{q^2}{2}\left[ \mp \frac{\mu'}{2\mu_0}f_{1,0}-1 \right] \;.
\eeq
We can then conclude that for longitudinal perturbations ($\theta_{\bf q}=0$, negative sign in front of $\mu'$),
the homogenous solution is stable confirming the results of numerical analysis. In the case of transversal perturbations ($\theta_{\bf q}=\pm \pi/2$), the stability condition is given by
\beq
\mu_0 > \frac{\mu'^2}{4\xi}
\label{transverse-stab}
\eeq
meaning that close to the instability threshold of the disordered solution,
when $\mu_0$ is positive but small, the state of homogeneous order is
unstable with respect to long wavelength perturbations. This instability
was first identified in a kinetic-equation analysis by Shi and Ma
\cite{ShiMa}.
Note that condition (\ref{transverse-stab}) is valid up to the third order
in $\epsilon$ (or, equivalently, in the order parameter $||{\bf w}||$).
It yields the stability line
\beq
\rho_s = \frac{4\mu_2-\mu'^2\xi_2}{\mu'^2\xi_1-4\mu'}\,,
\eeq
where $\mu_2=\mu(\rho=0)$, $\xi_1=(1/\xi)'$ and $\xi_2=(1/\xi(\rho_0=0))$.
We do not provide here the explicit analytical expression for $\sigma_s$
because this requires solving a sixth order polynomial.

We remark that the near-threshold instability discussed above is rather generic and appears in
``dry'' active matter systems with metric interactions, as opposed to systems with metric-free ones, 
where the interaction rate is density-independent, and $\mu'=0$
\cite{Gopinath,Peshkov2012a,Ihle2012}. In this case (topological active
nematics), stability would be enforced by the positive higher order
corrections $\mu_0 q^2$ which dominates arbitrarily close to
threshold.

\section{Inhomogeneous solution}

We now show how a spatially-inhomogeneous stationary ``band" solution to our hydrodynamic 
equations can be found. First we remark that our equation for the nematic field Eq.~(\ref{eq:last}) is formally
the same as that derived in \cite{Peshkov2012b} for polar particles
with nematic alignment when the polar field is set to zero, as it is imposed here by the complete nematic symmetry of our system.
We thus expect an ordered band solution made of two fronts connecting a linearly stable homogeneous disordered state ($\rho=\rho_{\rm gas}< \rho_t$) and a linearly stable homogeneous ordered state
 ($\rho=\rho_{\rm band}> \rho_s$) (see Fig.~\ref{fig:stab_diag}). Following \cite{Peshkov2012b}, we rewrite 
\beq
\mu(\rho)={\mu'}(\rho-\rho_{\rm t})\,,
\eeq
with $\rho_{\rm t} = (1-\hat P_1)/{\mu'}$, suppose that the nematic field is aligned along one of the axes and varies only
along $y$. In other words:
\beq
{\rm Re}\left(f_{1}\right)=f_1(y) \;\;,\;\;\  {\rm Im}\left(f_{1}\right)=0 \;\;,\;\;\rho=\rho(y)\,.
\label{ref-frame2}
\eeq
Eqs.~(\ref{eq:rho2}) then becomes 
\beq
\p_y^2 \rho = \p_y^2 f_1
\label{eq:band1}
\eeq
which can be integrated to give 
\beq
\rho= f_1+ A y + \rho_{\rm gas}
\label{eq:band2}
\eeq
where $A$ and $ \rho_{\rm gas}$ are integration constants. Furthermore, to keep
the fields finite for $|y| \to \infty$, one has $A=0$. By substituting
Eqs.~(\ref{eq:band1}) and (\ref{eq:band2}) into Eq.~(\ref{eq:last}) one gets 
\beq
\partial_{yy}f_{1}=-4{\mu'}\left(\rho_{\rm gas}-\rho_{\rm t}\right)f_{1}-4{\mu'}f_{1}^{2}+4\xi f_{1}^{3}
\label{eq:band3}
\eeq
We multiply equation (\ref{eq:band3}) by $\partial_y f_1$ and integrate it once to obtain
\beq
\frac{1}{2} \left( \partial_{y} f_{1}\right)^2 = -2{\mu'}\left(\rho_{\rm gas}-\rho_{\rm t}\right)f_{1}^{2}-\frac{4}{3}{\mu'}f_{1}^{3}+\xi f_{1}^{4}.
\eeq
Separating the variables we obtain  
\beq
\int dy=\pm \int\frac{df_{1}}{\sqrt{-4{\mu'}\left(\rho_{\rm gas}-\rho_{\rm t}\right)f_{1}^{2}-\frac{8}{3}{\mu'}f_{1}^{3}+2\xi f_{1}^{4}}}
\eeq
Integration of this equation under the condition  ${\lim}_{y\rightarrow\pm\infty}f_{1}(y){=}0$
gives after simplifications
\beq
f_1\left(y\right)=\frac{3\left(\rho_{\rm t}-\rho_{\rm gas}\right)}{1+a\cosh\left(2y\sqrt{{\mu'}\left(\rho_{\rm t}-\rho_{\rm gas}\right)}\right)}
\label{eq:f1_solution}
\eeq
where $a=\sqrt{1-\frac{9\xi}{2{\mu'}}\left(\rho_{\rm t}-\rho_{\rm gas}\right)}$.
We still need to obtain the value of $\rho_{\rm gas}$ which is fixed by the condition 
$\int_{L}\rho\left(y\right)dy=\rho_{0}L$, where $L$ is the length of the box.
In the integral on the l.h.s we can neglect the exponentially decaying tails and integrate
instead on the infinite domain. Furthermore, in the limit $L\rightarrow \infty$ we can
neglect the exponentially weak dependence of $\rho_{\rm gas}$ on $L$ everywhere except 
the $a$ term. We then obtain
\begin{eqnarray}
\rho_{\rm gas} & \approx & \rho_{\rm t}-\frac{2{\mu'}}{9\xi}\left(1-4e^{-KL}\right) \;,\\
K & = & \frac{2\sqrt{2} {\mu'}}{9\sqrt{\xi}}\left(1+\frac{9\xi}{2{\mu'}}\left(\rho_{0}-\rho_{\rm t}\right)\right) \;.
\label{eq:rho_gas}
\end{eqnarray}
Substituting it back into Eq.~(\ref{eq:f1_solution}) we get, under the assumption $L\rightarrow \infty$:
\beq
\label{eq:f1-band}
f_1\left(y\right)=\frac{f_{1}^{\rm band}}{\left(1+2e^{-\frac{KL}{2}}\cosh\left(y\frac{2\sqrt{2}{\mu'}}{3\sqrt{\xi}}\right)\right)}
\;\;{\rm where} \;\; 
f_{1}^{\rm band}=\frac{2{\mu'}}{3\xi}
\eeq
and we finally obtain the ordered solution density
\beq
\rho_{\rm band} = f_{1}^{\rm band}  + \rho_{\rm gas} =  \rho_{\rm t}+\frac{4{\mu'}}{9\xi}\left(1+2e^{-KL}\right)
\eeq
with, as expected, $\rho_{\rm band}>\rho_t>\rho_{\rm gas}$, which guarantees the
stability of both the ordered and disordered parts of the solution.
Note that since $f_1^{\rm band}>0$ the nematic order is parallel to the
$x$ direction (i.e. along the band orientation).
This is the opposite of what happens in
the Vicsek model, where bands extend transversally with respect to their polarization
\cite{Chate1}.

We can introduce the band fraction $\Omega$ which indicates the fraction of the box occupied by the band.
If we suppose that the front width is negligible (once again justified in the limit $L\rightarrow \infty$), this band fraction is determined by the equation
\beq
\Omega\left(\rho_{\rm band}-\rho_{\rm gas}\right)+\rho_{\rm gas}  =  \rho_{0}
\eeq
Substituting inside the values of $\rho_{\rm gas}$ and $\rho_{\rm band}$, we obtain
\beq
\Omega=\frac{9\xi\left(\rho_{0}-\rho_{\rm t}\right)+2{\mu'}}{6{\mu'}}
\label{eq:omega_eq}
\eeq
The condition $0<\Omega<1$ gives us the lower $\sigma_{\rm min}$ and upper $\sigma_{\rm max}$ 
limits of the existence of bands.
As found for polar particles aligning nematically, these limits of existence of the band solution extend 
{\it beyond} the region of linear instability of the homogeneous ordered solution 
(given by $\sigma\in [\sigma_{\rm s}, \sigma_{\rm t}]$, see Fig.~\ref{fig:stab_diag}). 
In Fig.~\ref{fig:stab_diag}, we provide a graphical illustration of the shape and properties of the band solution.

An important problem left for future work is the linear stability analysis of the band solution in two space dimensions.
This is all the more important as the unpublished work of Shi and Ma \cite{ShiMa} suggests the existence of some instability mechanism.

\section{Langevin formulation}
\label{sec-langevin}

Being based on a master equation, the derivation we have discussed in the previous
sections leads to a set of deterministic PDEs. This is a standard approach in equilibrium
statistical physics, where the microscopic fluctuations are integrated out in the coarse graining process
implicit in the definition of a mesoscopic cell size $\ell_B$. Fluctuations, when needed, can be eventually introduced
as an additive, delta correlated stochastic term as in Ref. \cite{TT}. However, the presence of large density
fluctuations \cite{Ramaswamy2003} suggests that fluctuations may not be faithfully accounted for by some additive noise
term. The precise nature of noise correlations at
the mesoscopic level cannot be safely overlooked in non-equilibrium systems, as it is known
that stochastic terms multiplicative in the relevant fields can radically alter the universality class of mesoscopic theories \cite{Munoz}.

In this section, we perform a direct coarse-graining of the microscopic dynamics in order to
compute the (multiplicative) stochastic terms which emerge at the mesoscopic level.
We however restrict the computation to the stochastic terms emerging from the collisionless dynamics.
%
For real-space coarse-graining, we make use of a smooth, isotropic, normalized 
(to one) filter $g_s(r)$ decaying exponentially 
or faster for $r>s$, e.g., a Gaussian of width $s$.
The fluctuating coarse-grained density and nematic order field are then defined as
\beq
\rho(\bsf{x}, t)  \equiv \sum_{i=1}^N g_s({\bsf x}_i^t-\bsf{x})
\label{cg1}
\eeq
and
\beq
{\bf w}(\bsf{x}, t) \equiv 
 \sum_{i=1}^N  g_s({\bsf x}_i^t-\bsf{x}) \tilde{\bf Q}_i^t \;,
\label{cg2}
\eeq
where we have introduced the microscopic
traceless tensor
\begin{equation}
\tilde{\bf Q}^t_i = \hat{\bf n}^t_i \hat{\bf n}^t_i - \frac{\mathbb{I}}{2} = \frac{1}{2} \left(\begin{array}{cc}
\cos 2\theta^t_i & \sin 2\theta^t_i\\
\sin 2\theta^t_i & -\cos 2\theta^t_i \end{array} \right) \equiv {\bf Q}(\theta_i^t) \;.
\label{Qmicro}
\end{equation}

\subsection{Density field fluctuations}
The correlations of density field fluctuations can be derived
by generalizing an approach first outlined by Dean \cite{Dean} for Brownian particles. As mentioned above, we use the collisionless dynamics.
We are interested in the time evolution of the density field (\ref{cg1}), which
is given by 
\beq
\rho(\bsf{x}, t+\Delta t) =  \sum_{i=1}^N g_s({\bsf x}_i^{t+\Delta
  t}-\bsf{x}) = \sum_{i=1}^N g_s({\bsf x}_i^t+\Delta {\bsf x}_i^t -\bsf{x}) 
\label{eql1}
\eeq
where $\Delta \bsf{x}_i^t= \bsf{x}_i^{t+\Delta t}-\bsf{x}_i^t$.\\
Expanding up to second order in powers of  
$\Delta \bsf{x}_i^t$ according to It$\hat{\rm o}$ calculus \cite{Ito} 
and by virtue of Eq.~(\ref{eq:stream}) one has
\beq
\p_t \rho(\bsf{x}, t) = T_0(\bsf{x}, t) + T_1(\bsf{x}, t) \,,
\eeq
where
\begin{equation}
T_0(\bsf{x}, t) = \frac{d_{\rm 0}^2}{2 \tau_d}
\sum_{i=1}^N [\hat{\bf n}_i^t]_\alpha [\hat{\bf n}_i^t]_\beta\p_\alpha \p_\beta g_s(\bsf{x}_i^t-\bsf{x})
\end{equation}
and
\begin{equation} 
T_1(\bsf{x}, t) = \frac{d_0}{\tau_d} \sum_{i=1}^N \kappa_i^t \, \left(\hat{\bf n}_i^t \cdot \nabla\right) 
g_s(\bsf{x}_i^t-\bsf{x})\,.
\end{equation}
Note that derivatives are taken w.r.t.~the argument of the function $g_s$,
and not w.r.t.~$\bsf{x}$.
The second order term $T_0$ yields the deterministic part of the density dynamics. 
By Eqs.~(\ref{cg1},\ref{cg2}) and the definition of the microscopic
nematic tensor $\tilde{\bf Q}$ [Eq.~(\ref{Qmicro})] one easily gets 
\beq
T_0 = \frac{D_0}{2} ({\bf \Gamma} : {\bf w} ) + \frac{D_0}{2} \nabla^2 \rho\,,
\eeq
that is, the right hand side of the diffusion Eq.~(\ref{eq:rho2}).
The first-order term $T_1$ gives rise to the (zero average) stochastic term we are
interested in. At this stage, $T_1$ is not a simple function of the mesoscopic fields; however, following Ref.~\cite{Dean} 
it is possible to show that its two point correlation can be recast
as a function of $\rho$ and ${\bf w}$.
Averaging over the random numbers $\kappa_i^t$,
we have, in the limit $s \to 0$,
\begin{eqnarray}
\hspace{-1 cm}
\label{eq:corr0}
\nonumber
\langle T_1(\bsf{x}, t) T_1(\bsf{y}, t') \rangle &=& d_0^2 \frac{\delta(t-t')}{\tau_d} \sum_{i=1}^N \left(\hat{\bf n}_i^t \cdot \nabla_x\right) \left(\hat{\bf n}_i^t \cdot \nabla_y\right) g_s(\bsf{x}_i^t-\bsf{x}) g_s(\bsf{x}_i^t-\bsf{y})\\
&\simeq& d_0^2 \frac{\delta(t-t')}{\tau_d} \sum_{i=1}^N \left(\hat{\bf n}_i^t \cdot \nabla_x\right) \left(\hat{\bf n}_i^t \cdot \nabla_y\right) \Big( g_s (\bsf{x}\!-\!\bsf{y}) g_s(\bsf{x}_i^t\!-\!\bsf{x}) \Big) \;.
\end{eqnarray}
Using Eq.~(\ref{Qmicro}), one then finds, approximating the filter $g_s$ by a
Dirac delta in the limit $s \to 0$,
\begin{equation}
\hspace{-1 cm}
\langle T_1(\bsf{x}, t) T_1(\bsf{y}, t') \rangle = d_0^2 \frac{\delta(t-t')}{\tau_d} \partial_{\alpha} \partial_{\beta} \left[ \delta(\bsf{x}\!-\!\bsf{y}) 
\left( w_{\alpha \beta}(\bsf{x}, t) + \frac{1}{2} \rho(\bsf{x}, t) \delta_{\alpha \beta} \right) \right]
\end{equation}
We can rewrite the noise term $T_1$ in the stochastically equivalent (i.e., with the same correlations on the mesoscopic scale) form 
\beq
T_1(\bsf{x}, t) = \nabla \cdot {\bf h}(\bsf{x}, t)
\eeq
where ${\bf h}$ is a Gaussian, zero-average vectorial noise, delta-correlated in time with correlations
\begin{equation}
\langle h_{\alpha}(\bsf{x}, t) h_{\beta}(\bsf{y}, t') \rangle 
\simeq \frac{d_0^2}{\tau_d}\, \delta(t-t')\, \delta (\bsf{x}-\bsf{y})
\!\left( w_{\alpha \beta}(\bsf{x}, t)
  + \frac{\delta_{\alpha\!\beta}}{2} \rho (\bsf{x}, t)\right) .
\label{eq:corr}
\end{equation}
Such a noise term can finally be expressed in the more convenient form  
\beq
h_{\alpha}(\bsf{x}, t) = K_{\alpha \beta}(\bsf{x}, t) {\tilde h}_{\beta}(\bsf{x}, t)\,, 
\eeq
where the Gaussian noise 
${\bf {\tilde h}}$ has correlations independent from the hydrodynamic fields
\beq
\langle {\tilde h}_{\alpha}(\bsf{x}, t)
{\tilde h}_{\beta}(\bsf{x}', t') \rangle
= 2 D_0\, \delta_{\alpha\beta}\, \delta(t-t')\, \delta(\bsf{x}-\bsf{x}')
\eeq 
and
the tensor ${\bf K}$ is implicitly defined from the relation 
${\bf K}\cdot {\bf K}=(\rho /2) {\bf I}+{\bf w}$ 
(with ${\bf I}$ being the identity matrix).
In the limit of small ${\bf w}$ considered here, we can expand
${\bf K}$ to first order in ${\bf w}$, yielding
\beq
\label{eq:defK}
{\bf K} = \frac{1}{\sqrt{2}}\, \rho^{1/2} \left( {\bf I}+\frac{{\bf w}}{\rho} \right).
\eeq 
The divergence term $\nabla\cdot$ appearing in $T_1$ reflects global density conservation, while the proportionality of noise variance to number density 
can be interpreted as a consequence of the central limit theorem.
Adding up the two contributions, one finally gets
\begin{equation}
\p_t \rho = \frac{D_0}{2} ({\bf \Gamma} : {\bf w} ) + \frac{D_0}{2}
\nabla^2 \rho + \nabla\cdot \,({\bf K}\cdot {\bf {\tilde h}})\,.
\label{eq:rho_stoc}
\end{equation}

\subsection{Nematic field fluctuations}

We next discuss fluctuations of the nematic tensor.
As seen from Eq.~(\ref{cg2}), ${\bf w}$ is a function of the $2 N$ microscopic stochastic
variables $\bsf{x}_i^t$ and ---through the microscopic nematic tensor
(\ref{Qmicro})--- $\theta_i^t$, whose dynamics is given by
Eqs.~(\ref{eq:align})-(\ref{eq:stream}). 
According to It\^o calculus, one has
\beq
\p_t {\bf w} = {\bf \Omega}_0 + {\bf \Omega}_1 + {\bf \Omega}_2
\eeq
where ${\bf \Omega}_0$ is the deterministic part of the coarse-grained
collisionless dynamics (which we do not write here explicitly),
while ${\bf \Omega}_1$ and ${\bf \Omega}_2$ are two stochastic contributions,
\begin{eqnarray}
{\bf \Omega}_1 &=& \frac{2}{\tau_d} \sum_{i=1}^N g_s(\bsf{x}_i^t-\bsf{x})\, {\bf
  A}\cdot \tilde{\bf Q}_i^t \,\psi_i^t \\
{\bf \Omega}_2 &=& \frac{d_0}{\tau_d} \sum_{i=1}^N \kappa_i^t \hat{\bf n}_i^t \cdot \nabla g_s(\bsf{x}_i^t-\bsf{x})\, \tilde{\bf Q}_i^t 
\end{eqnarray}
where $\psi^t_i$ and $\kappa_i^t$ are the microscopic noises and
\begin{equation}
{\bf A} = \left(\begin{array}{cc}
0 & -1\\
1 & 0 \end{array} \right) \;.
\label{eq:A}
\end{equation}
Note that in ${\bf \Omega}_1$ we have retained only the linear contribution
in the microscopic noise $\psi_i^t$.
We first focus on the stochastic terms ${\bf \Omega}_1$.
On coarse-graining scales, averaging over the microscopic noise $\psi_i^t$,
correlations of ${\bf \Omega}_1$ are given by
\begin{eqnarray}
\label{eq:corrQ}
\nonumber
\hspace{-2.5cm}\langle \left[ {\bf \Omega}_1(\bsf{x}, t)\right]_{\alpha \beta} \left[{\bf \Omega}_1(\bsf{y}, t')\right]_{\gamma \delta} \rangle
&=& 4 \eta^2 \frac{\delta(t-t')}{\tau_d} \sum_{i=1}^N 
g_s(\bsf{x}_i^t-\bsf{x}) g_s(\bsf{x}_i^t-\bsf{y}) \left[{\bf A}\cdot
  \tilde{\bf Q}_i^t \right]_{\alpha \beta}\,\left[{\bf A}\cdot \tilde{\bf Q}_i^t \right]_{\gamma \delta}\\
&\approx& 4\eta^2 \frac{\delta(t-t')}{\tau_d} g_s(\bsf{y}\!-\!\bsf{x})
\sum_{i=1}^N g_s(\bsf{x}_i^t\!-\!\bsf{x}) 
\left[{\bf A} \cdot \tilde{\bf Q}_i^t \right]_{\alpha \beta} \,
\left[{\bf A} \cdot \tilde{\bf Q}_i^t \right]_{\gamma \delta}
\end{eqnarray}
To evaluate this correlator, we determine the average value
$\langle \sum_i g_s\, ({\bf A} \cdot \tilde{\bf Q}_i^t) ({\bf A} \cdot \tilde{\bf Q}_i^t) \rangle$, in the framework of the deterministic dynamics studied in Sect.~\ref{sect-kinetic-theory}, namely
\begin{equation}
\hspace{-2.3cm} \left< \sum_{i=1}^N g_s(\bsf{x}_i^t\!-\!\bsf{x}) \left[{\bf A} \cdot \tilde{\bf Q}_i^t\right]_{\alpha \beta} \left[{\bf A} \cdot \tilde{\bf Q}_i^t\right]_{\gamma \delta} \right> 
= \int_{-\frac{\pi}{2}}^{\frac{\pi}{2}} d\theta\, f(\bsf{x},\theta,t)\left[{\bf A} \cdot {\bf Q}(\theta)\right]_{\alpha \beta} \left[{\bf A} \cdot {\bf Q}(\theta) \right]_{\gamma \delta}
\end{equation}
After some rather lengthy calculations, using the closure equations (\ref{eq-f2-closure}), one finds
\begin{eqnarray}
\nonumber
\hspace{-2.4cm} \int_{-\frac{\pi}{2}}^{\frac{\pi}{2}} d\theta\, f(\bsf{x},\theta,t)\left[{\bf A} \cdot {\bf Q}(\theta)\right]_{\alpha \beta} \left[{\bf A} \cdot {\bf Q}(\theta)\right]_{\gamma \delta} &=& \rho \, J_{\alpha \beta \gamma \delta} + \frac{2b_2}{a_2} \left[ (w_{\mu \nu} w_{\mu \nu}) J_{\alpha \beta \gamma \delta} - 2 w_{\alpha \beta} w_{\gamma \delta} \right]\\
&+& \frac{1}{4a_2} \left[ \Gamma_{\mu \nu} w_{\mu \nu}
J_{\alpha \beta \gamma \delta} - \Gamma_{\alpha \beta} w_{\gamma \delta} - \Gamma_{\gamma \delta} w_{\alpha \beta} \right],
\label{eq-AQAQ}
\end{eqnarray}
where we have introduced the tensor
\beq
J_{\alpha \beta \gamma \delta} = \frac{1}{2} \left( \delta_{\alpha \gamma}
\delta_{\beta \delta} + \delta_{\alpha \delta} \delta_{\beta \gamma}
- \delta_{\alpha \beta} \delta_{\gamma \delta} \right)
\eeq
which plays the role of a unit tensor for the double contraction of symmetric traceless tensors,
e.g., $w_{\alpha \beta} = J_{\alpha \beta \mu \nu}\, w_{\mu \nu}$.
In order to characterize the noise ${\bf \Omega}_1$, we introduce
the following change of variables:
\beq
\left[{\bf \Omega}_1(\bsf{x}, t)\right]_{\alpha \beta} = H_{\alpha \beta \mu \nu}(\bsf{x}, t) \,
\tilde{\Omega}_{\mu \nu}(\bsf{x}, t)
\eeq
where ${\bf \tilde{\Omega}}$ is a tensorial symmetric traceless white noise,
such that
\beq
\langle \tilde{\Omega}_{\alpha \beta}(\bsf{x}, t) \tilde{\Omega}_{\gamma \delta}(\bsf{y}, t') \rangle
= 2 D \delta(\bsf{x}-\bsf{y})\, \delta(t-t') \, J_{\alpha \beta \gamma \delta},
\eeq
with $D=2\eta^2/\tau_d$.
The correlation of ${\bf \Omega}_1$ then reads
\beq \label{eq:corrQ2}
\langle \left[ {\bf \Omega}_1(\bsf{x}, t)\right]_{\alpha \beta} \left[{\bf \Omega}_1(\bsf{y}, t')\right]_{\gamma \delta} \rangle
= 2 D \delta(\bsf{x}-\bsf{y})\, \delta(t-t') \, H_{\alpha \beta \mu \nu}(\bsf{x}, t) H_{\gamma \delta \mu \nu}(\bsf{x}, t)
\eeq
By identification with Eq.~(\ref{eq:corrQ}), and using Eq.~(\ref{eq-AQAQ}),
one eventually finds for ${\bf H}$
\begin{eqnarray}
\nonumber
H_{\alpha \beta \gamma \delta} &=& \rho^{1/2}\, J_{\alpha \beta \gamma \delta}
+\frac{b_2}{a_2\, \rho^{1/2}} \left[ w_{\mu \nu} w_{\mu \nu} J_{\alpha \beta \gamma \delta} - 2 w_{\alpha \beta} w_{\gamma \delta} \right]\\
&& \qquad \qquad \quad + \frac{1}{8a_2\, \rho^{1/2}} \left[ \Gamma_{\mu \nu} w_{\mu \nu} J_{\alpha \beta \gamma \delta} - \Gamma_{\alpha \beta} w_{\gamma \delta} - \Gamma_{\gamma \delta} w_{\alpha \beta} \right].
\label{eq:defH}
\end{eqnarray}
Note that, in agreement with the central limit theorem, ${\bf \Omega}_1$ is
(at least to first order in ${\bf w}$)
proportional to the square root of local density.

The second stochastic term ${\bf \Omega}_2$, finally, can be treated similarly,
but it would give rise to a conserved noise (due to the presence of
$\nabla$ terms) akin to the one discussed for the density equations,
thus related to density fluctuations affecting the ${\bf w}=\rho{\bf Q}$ field.
We discard such conserved term as irrelevant (in the renormalization group sense) with respect
to the non-conserved multiplicative noise ${\bf \Omega}_1$.

In order to write down the complete Langevin equation, one also needs to evaluate
the contribution of the deterministic part ${\bf \Omega}_0$.
However, expressing this contribution in terms of the fluctuating fields $\rho$
and ${\bf w}$ turns out to be a very complicated task. One should also take into account collisions between particles, and not only the collisionless dynamics described
by ${\bf \Omega}_0$. Then some further approximations would be required to treat
the non-linear part of the dynamics.

In addition, microscopic collisions could provide a further fluctuation source due to disorder below the coarse-graining scale. While we conjecture them to be irrelevant, we leave a final settlement of this difficult problem for future work, and use for the deterministic part of the dynamics the terms the hydrodynamic equation (\ref{eq:Q2}), derived from the Boltzmann approach.

We thus finally obtain the stochastic equation for the nematic field
\begin{equation}
\p_t {\bf w} = \mu \,{\bf w} - 2\xi  \,{\bf w} \left({\bf w}:{\bf w}\right) 
+ \frac{1}{2} \Delta {\bf w}  + \frac{1}{8}{\bf \Gamma} \rho
+ {\bf H}:{\bf \tilde{\Omega}} \;.
\label{eq:QS}
\end{equation}
A few remarks are in order: first, our expressions of the noise amplitudes ${\bf K}$ and ${\bf H}$ (Eqs.(\ref{eq:defK}) and (\ref{eq:defH})) suggest that the stochastic terms might be better expressed in terms of the field ${\bf  Q}$, rather than ${\bf w}=\rho{\bf Q}$; second, Eqs. (\ref{eq:rho_stoc}) and (\ref{eq:QS}) are also derived from
an apolar Vicsek-style model in \cite{shradha_thesis}.

In spite of the limitations listed above, the present approach already provides us with useful information on the statistics of the noise terms, which is seen to differ significantly from the white noise postulated on a phenomenological basis in previous works. On top of the overal $\rho^{1/2}$ dependency, our calculation reveals a non-trivial dependence of the correlation of the noise on the nematic order parameter
[see Eqs.~(\ref{eq:defK}, \ref{eq:rho_stoc}, \ref{eq:corrQ2}, \ref{eq:defH})].

\section{Conclusions}

To summarize, using as a starting point the simple active nematics model of \cite{Chate2}, 
we have demonstrated how one can derive in a systematic manner a continuous mesoscopic description:
We formulated a version of the Boltzmann-Ginzburg-Landau approach put forward in \cite{Peshkov2012a,Peshkov2012b} for this case where
(anisotropic) diffusion dominates, deriving a simple hydrodynamic equation for the nematic ordering field --Eq.~(\ref{eq:Q2}). 
We have then used a direct coarse-graining approach to endow the hydrodynamic equations with proper noise terms. 

The next stage, left for future work, consists in studying the stochastic PDEs obtained.
At the linear level, it is clear that in the long wavelength limit, standard results on giant density fluctuations \cite{Ramaswamy2003} are recovered. However, the large amplitude of density fluctuations calls for a non-linear analysis (which turns out to be very difficult), where the density dependence of the noise derived in Sect.~\ref{sec-langevin} may play an important role. Ideally, one should try to tackle this issue by applying methods from field theory and renormalization group analysis. 
In addition, we note that the multiplicative nature of the noise may also affect finite-wavelength properties, like coarsening behavior.
The analysis of the stochastic PDEs can be done numerically,
but some care must be taken when dealing with the multiplicative, conserved noise terms in (\ref{eq:rho_stoc}). 

Pending such attempts, some remarks and comments are already in order: like all previous cases studied before, 
the hydrodynamic equations found exhibit a domain of linear instability of the homogeneous ordered solution bordering
the basic transition line $\sigma_{\rm t}$. This solution does become linearly stable deeper in the ordered phase 
(for $\sigma$ below $\sigma_{\rm s}$).
Moreover, we have found that the long wavelength instability of the homogeneous ordered solution leads to a nonlinear, inhomogeneous band solution
--see Eq.~(\ref{eq:f1-band})-- and that this band solution exists beyond the 
$[\sigma_{\rm s},\sigma_{\rm t}]$ interval. These coexistence regions suggest, at the fluctuating level, 
{\it discontinuous} transitions. 

This seems to be at odds with the reported behavior of the original microscopic
model: (i) the order/disorder transition has been reported to be of the
Kosterlitz-Thouless type \cite{Chate2}; (ii) there is no trace, at the
microscopic level, of the existence of a non-segregated, homogeneous phase;
(iii) coming back to giant number fluctuations, we note that the standard
calculation is made in the homogeneous ordered phase whereas the numerical
evidence for them reported in \cite{Chate2} appears now to have been obtained in
the inhomogeneous phase. All this calls for revisiting the simple particle-based
model and, eventually, understanding its behaviour in the context of
the stochastic continuum theory constructed here.

\subsection*{Acknowledgements}

Part of this work was performed at the Max Planck
Institute for the Physics of Complex Systems in Dresden,
Germany, within the Advanced Study Group 2011/2012
'Statistical Physics of Collective Motion'. F.G. acknowledges
support by EPSRC First Grant EP/K018450/1.

\appendix
\section{Fourier expansion of the master equation}
\label{App-Fourier-MEq}

We provide in this Appendix details of the Fourier expansion of the master equation (\ref{B2}), leading to Eq.~(\ref{Hydro0}).
Multiplying Eq.~(\ref{B2}) by $e^{i 2 \theta}$ and integrating over $\theta$, one gets
\begin{eqnarray}
\nonumber
\partial_t \hat{f}_k &=&
\partial_{\alpha}\partial_{\beta} \int_{-\pi/2}^{\pi/2} d \theta\, e^{i 2 k \theta}
\hat{n}_{\alpha}(\theta) \hat{n}_{\beta}(\theta) f(\bsf{x}, \theta, t)\\
&& \qquad \qquad + \int_{-\pi/2}^{\pi/2} d \theta\, e^{i 2 k \theta} I_{\rm diff}[f]
+ \int_{-\pi/2}^{\pi/2} d \theta\, e^{i 2 k \theta} I_{\rm coll}[f,f]\,.
\label{MEqFourier}
\end{eqnarray}
In the following, we successively compute each term of the r.h.s.~of Eq.~(\ref{MEqFourier}).

\subsection{Diffusion-like term}
Let us define $Q_{\alpha \beta}(\theta)$ as
\beq
Q_{\alpha \beta}(\theta) = 
\hat{n}_{\alpha}(\theta) \hat{n}_{\beta}(\theta)-\frac{\delta_{\alpha \beta}}{2}.
\eeq
We then have
\begin{eqnarray}
Q_{11}(\theta, t) = -Q_{22}(\theta, t) &=& \frac{1}{2} \cos 2\theta =
\frac{e^{i 2 \theta}+e^{-i 2 \theta}}{4} \;,\nonumber \\
Q_{12}(\theta, t) = Q_{21}(\theta, t) &=& \frac{1}{2} \sin 2\theta =
\frac{e^{i 2 \theta}-e^{-i 2 \theta}}{4 i} \;.
\end{eqnarray}
As a result,
\begin{eqnarray}
\hspace{-1 cm}
\p_\alpha \p_\beta \int_{-\pi/2}^{\pi/2} d \theta\, e^{i 2 k \theta} 
\hat{n}_{\alpha}(\theta) \hat{n}_{\beta}(\theta) f(\theta)&=&
 \p_\alpha \p_\beta \int_{-\pi/2}^{\pi/2} d \theta e^{i 2 k \theta} 
\left(Q_{\alpha\beta}(\theta)+\frac{\delta_{\alpha\beta}}{2}\right) f(\theta) \nonumber \\
&=& \frac{1}{2} \Delta \hat{f}_k 
+  \frac{1}{4} \left( \nabla^{*2} \hat{f}_{k+1} + \nabla^{2} \hat{f}_{k-1} \right)
\end{eqnarray}

\subsection{Self-diffusion term}

We have rather straightforwardly
\begin{eqnarray}
\int_{-\pi/2}^{\pi/2} d \theta\, e^{i 2 k \theta} I_{\rm diff}[f] &=& - \hat{f}_{k}
+ \int_{-\pi/2}^{\pi/2} d \theta' e^{i 2 k \theta'} f(\theta') 
\int_{-\infty}^{\infty} d \zeta
e^{i 2 k \zeta } P(\zeta)\nonumber\\
&=& \left[ \hat{P}_k -1 \right]\hat{f}_k
\end{eqnarray}
where
\beq
\hat{P}_k=\int_{-\infty}^{\infty} d \zeta\, e^{i 2 k \zeta } P(\zeta)\,.
\eeq
is the Fourier transform of $P(\zeta)$.

\subsection{Binary collisions term}

Let us split the Fourier transformed collision integral
into an outgoing (negative) collision term $I_k^{(-)}$ 
and an ingoing (positive) collision term $I_k^{(+)}$.
A direct integration of the outgoing collision term yields,
using $K(\theta,\theta')=\tilde{K}(\theta-\theta')$,
\beq
I_k^{(-)} \equiv -\int_{-\pi/2}^{\pi/2} d \theta\, e^{i 2 k \theta} 
f(\theta) \int_{-\pi/2}^{\pi/2} d \theta' f(\theta') \tilde{K}(\theta-\theta')
=-\frac{1}{\pi}\sum_q \hat{K}_q \hat{f}_q \hat{f}_{k-q}
\eeq
where $\hat{K}_q$ is the Fourier coefficient of $\tilde{K}(\theta-\theta')$
given by, using Eq.~(\ref{collisionkernel}),
\beq
\hat{K}_q = \int_{-\pi/2}^{\pi/2} d \theta 
 e^{i 2 q \theta} \left[ \left|\sin \frac{\theta - \theta'}{2}\right|+
\left|\cos \frac{\theta - \theta'}{2}\right|\right] = 
\frac{4}{1-16 q^2} \;.
\eeq
Then, the calculation of the ingoing collision term requires a few steps.
After integration of the (generalized) Dirac delta $\delta_{\pi}$, we have
\begin{equation}
I_k^{(+)} = \hat{P}_k \int_{-\pi/2}^{\pi/2} d \theta_1 \int_{-\pi/2}^{\pi/2} d \theta_2
\, e^{i 2k \Psi(\theta_1,\theta_2)}
f(\theta_1) \tilde{K}(\theta_1\!-\!\theta_2)  f(\theta_2) \;.
\label{coll2}
\end{equation}
By the change of variables $\phi=\theta_1-\theta_2$, one gets
\begin{equation}
I_k^{(+)} = \hat{P}_k \int_{-\pi/2}^{\pi/2} d \theta_2 \int_{-\pi/2-\theta_2}^{\pi/2-\theta_2}
d\phi \, e^{i 2k \Psi(\theta_2+\phi,\theta_2)} f(\theta_2+\phi) \tilde{K}(\phi) f(\theta_2) \;.
\end{equation}
Using the $\pi$-periodicity of the integrand with respect to $\phi$,
we can change the integration interval on $\phi$, yielding
\begin{equation}
I_k^{(+)} = \hat{P}_k \int_{-\pi/2}^{\pi/2} d \theta_2 \int_{-\pi/2}^{\pi/2}
d\phi \, e^{i 2k \Psi(\theta_2+\phi,\theta_2)} f(\theta_2+\phi) \tilde{K}(\phi) f(\theta_2)
\end{equation}
On this interval of $\phi$, one has from Eq.~(\ref{def:def-Phi})
\begin{equation}
\Psi(\theta_2+\phi,\theta_2)= \theta_2 + \frac{\phi}{2} \;.
\end{equation}
Expanding $f$ in Fourier series [see Eqs.~(\ref{expansion},\ref{expansion2})], we get
\begin{equation}
I_k^{(+)} = \frac{\hat{P}_k}{\pi^2} \sum_{q,q'} \hat{f}_q \hat{f}_{q'}
\int_{-\pi/2}^{\pi/2} d \theta_2 \, e^{i 2(k-q-q')\theta_2}
\int_{-\pi/2}^{\pi/2} d\phi \, e^{i (k-2q)\phi} \tilde{K}(\phi) \;.
\end{equation}
The integral over $\theta_2$ is equal to $\pi\delta_{k,q+q'}$.
Defining
\begin{equation}
\hat{J}_{k,q} = \int_{-\pi/2}^{\pi/2} d\phi \, e^{i (k-2q)\phi} \tilde{K}(\phi) \;,
\end{equation}
we finally obtain
\begin{equation}
I_k^{(+)} = \frac{\hat{P}_k}{\pi} \sum_{q} \hat{J}_{k,q} \hat{f}_q\hat{f}_{k-q} \;.
\end{equation}
The coefficient $\hat{J}_{k,q}$ can be computed explicitly, leading to
\begin{equation}
\hat{J}_{k,q} = 4 \, \frac{1+2\sqrt{2}(2q-k)(-1)^q\sin\left(\frac{k\pi}{2}\right)}{1-4(2q-k)^2}
\end{equation}
Note finally that $\hat{J}_{0,q}=\hat{K}_{q}$.

\section{Curvature-induced current and equilibrium limit} 
\label{app-curv-current}

In this Appendix, we show explicitly that the curvature-induced current,
that is the term $\frac{1}{2}{\rm Re}\left({\nabla^*}^{2}\hat{f}_{1}\right)$
appearing in the continuity equation (\ref{eq:RhoComplex}),
originates from the coupling of orientation with motility.
To this aim, we consider a
slightly generalized microscopic process w.r.t.~Eqs.~(\ref{eq:align},
\ref{eq:stream}), where particles are also allowed to move
perpendicular w.r.t to the nematic tensor. Replace
Eq.~(\ref{eq:stream}) by
\begin{equation}
\bsf{x}_i^{t+\Delta t}=\bsf{x}_i^{t} + d_{\rm 0}\,{\bf R}\left(\theta_i^t\right)
\label{eq:r}
\end{equation}
where ${\bf R}(\theta)$ is a stochastic operator defining the coupling
between orientation and particle motion, 
\begin{equation}
{\bf R}(\theta)=\left\{
\begin{array}{rl}
\hat{\bf n}(\theta) & \mathrm{w.p.}\;\;p/2\\
-\hat{\bf n}(\theta) & \mathrm{w.p.}\;\;p/2\\
\hat{\bf n}^{\perp}(\theta) & \mathrm{w.p.}\;\;(1-p)/2\\
-\hat{\bf n}^{\perp}(\theta) & \mathrm{w.p.}\;\;(1-p)/2
\end{array}\right.
\end{equation}
where $0 \leq p \leq 1$, w.p.~stands for ``with probability'' and
$\hat{\bf n}^{\perp}(\theta) = \hat{\bf n} (\theta+\pi/2)$ is the perpendicular
director. The standard active nematic case is recovered for $p=1$,
while $p=1/2$ corresponds to an isotropic random walk, a case for which motion is decorrelated from order.
The corresponding collisionless master equation reads
\begin{eqnarray}
\hspace{-1 cm}
f(\bsf{x}, \theta, t+\Delta t) &=& 
\frac{p}{2}\left[ f(\bsf{x}-\hat{\bf n}(\theta)  d_0 , \theta, t) 
+f(\bsf{x}+\hat{\bf n}(\theta)  d_0 , \theta, t)\right]\nonumber\\
&+&\frac{(1-p)}{2} \left[ f(\bsf{x}-\hat{\bf n}^{\perp}(\theta) d_0 , \theta, t) 
+f(\bsf{x}+\hat{\bf n}^{\perp}(\theta) d_0 , \theta, t) \right]\,.
\end{eqnarray}
By making use of It\^o calculus, one gets at the mesoscopic timescale $\tau_B$
\begin{equation}
\hspace{-1 cm}
\partial_t f(\bsf{x}, \theta, t) = 
 (2 p -1) \partial_{\alpha}\partial_{\beta}
\left[\hat{n}_{\alpha}(\theta) \hat{n}_\beta(\theta) \!-\! \frac{\delta_{\alpha
      \beta}}{2}\right]  f(\bsf{x}, \theta, t)
+ \frac{1}{2} \Delta  f(\bsf{x}, \theta, t)
\label{eq1}
\end{equation}
where  we have used the identity $\hat{n}^\perp_{\alpha}(\theta) \hat{n}^\perp_{\beta}(\theta) = \delta_{\alpha \beta} - \hat{n}_{\alpha}(\theta) \hat{n}_{\beta}(\theta)$. 
By considering the
zeroth-order Fourier term of $f$ (for which collision and angular
diffusion terms vanish), one obtains the continuity equation
\beq
\partial_{t}\rho =  \frac{1}{2}\Delta\rho + \frac{2p-1}{2}{\rm Re}
\left({\nabla^*}^{2}{f}_{1}\right)
\label{eq2}
\eeq
which shows that the non-equilibrium current vanishes for $p=\frac{1}{2}$.

\end{document}